\documentclass[useAMS,usenatbib,onecolumn]{mn2e}
\usepackage{times}

\usepackage{graphicx,color}

\newcommand{\beq}{\begin{equation}}
\newcommand{\beqa}{\begin{eqnarray}}
		  \newcommand{\eeq}{\end{equation}}
\newcommand{\eeqa}{\end{eqnarray}}

\newcommand{\lsim}{\la}
\newcommand{\gsim}{\ga}

\newcommand{\lmk}{\left(}
\newcommand{\rmk}{\right)}
\newcommand{\lnk}{\left\{ }
\newcommand{\rnk}{\right\} }

\newcommand{\lla}{\left\langle}

\newcommand{\rra}{\right\rangle}

\title[ Search for GWs from SBBHs with PTAs]{Search for  Memory and Inspiral Gravitational Waves from Super-Massive Binary Black Holes with Pulsar Timing Arrays} 
\author[N. Seto]{Naoki Seto$^{1}$
\\
$^{1}$Department of Physics, Kyoto University
Kyoto 606-8502, Japan
}

\begin{document}

\maketitle

%
%
%
%
%
%
\begin{abstract}
The merger of a super-massive binary black hole (SBBH) is one of the most extreme events in the universe with a huge amount of energy released by gravitational radiation. Although the characteristic gravitational wave (GW) frequency around the merger event is far higher than the nHz regime optimal for pulsar timing arrays (PTAs), nonlinear GW memory might be a critical smoking gun of the merger event detectable with PTAs. In this paper,  basic aspects of this interesting observation are discussed for SBBHs,  and  the detection numbers of their memory and inspiral GWs are estimated for ongoing and planned  PTAs.  We find  that the expected detection number would be smaller than unity for the two-types of signals even with the Square Kilometer Array. We also provide various scaling relations that would be useful to study detection probabilities  of GWs from individual SBBHs with PTAs.

\end{abstract}


\section{Introduction}

A pulsar is an  excellent clock in the universe, and provides us with a powerful method  to directly  detect gravitational waves (GWs) (Sazhin 1978; Detweiler 1979).  By  using multiple pulsars and suppressing noises due to independent timing fluctuations of individual pulsars,  we can  further improve sensitivity to GWs,  which generate common signals to  observed pulsars (Hellings 
\& Downs 1983). This statistical method is known as a pulsar timing array (PTA), and is now considered  as a promising approach to probe GWs around the nHz regime.
The most plausible target of ongoing PTAs is the stochastic GW background made by super-massive binary black holes (SBBHs).  
For detecting such a  background, a long term operation of a PTA is crucially advantageous.  Since the sensitivities of ongoing PTAs have been rapidly improved, it is likely that a PTA project would succeed in detecting the background before long (see {\it e.g.} Jenet et al. 2005; Hobbs et al. 2009).

Probing  GWs from individual SBBHs with a PTA is also quite interesting (Lommen 
\& Backer 2001; Sesana et al. 2009), especially in relation to traditional observations of electro-magnetic wave (EMW) emissions.  Actually, with a constraint on the GW amplitude, a pulsar timing observation (PSR B1855+09, Jenet et al. 2004) recently ruled out the previously proposed parameters of a SBBH system postulated from a radio observation of the galaxy 3C66B at $z=0.02$ (Sudou et al. 2003). It is expected that  interplays between  EMW observations and PTAs will be biased to objects at relatively low redshift, due to observational accessibilities (see {\it e.g.} Lommen 
\& Backer 2001).

On the evolutionary path of a SBBH, the most violent and fascinating phase would be its final merger where a  huge amount of energy is released by gravitational radiation. 
Impacts of such an extreme event would not be limited only to  the GW community. For example, it has been actively discussed that a transient EMW signature might be associated with a merger  of a SBBH ({\it e.g.} through interaction between a SBBH and its circumbinary disk, see Haiman et al. 2009 and references therein).  In  the future,  there might appear a potential  interpretation  that a peculiar time-dependent phenomena observed with EMWs is related to merger of a SBBH.

It would be very exciting to probe a merger event by observing the intense GWs with a PTA.
However,  the order of the characteristic frequency of the  merger GWs is given by $1/(2\pi M)\sim 3\times 10^{-5} (M/10^9 M_\odot)^{-1}$Hz ($M$: the total SBBH mass),  which is far higher than the optimal frequency regime of a PTA around $\sim$nHz.
Indeed, depending on the masses of SBBHs,  GWs around the merger phase are one of the primary targets of the proposed Laser Interferometer Space Antenna (LISA),  which is designed to  have sensitivity around 0.1mHz-1Hz (Bender et al. 1997). However,  LISA will not be launched  before 2019.

Fortunately, GW signals associated with the violent merger are not completely localized at the characteristic frequency $O(M^{-1})$. Because of  the intense and anisotropic GW emission around the merger phase, the so-called GW memory is simultaneously generated.  At frequencies much lower than $M^{-1}$, its waveform can be  regarded as a (burst-like) step function profile and its non-dimensional Fourier amplitude $h_c$ becomes  independent of the frequency.  Thus a GW memory might be a critical smoking gun of a SBBH merger event detectable  with a PTA.  If detected, it would surely have a  broad impact on astronomy.

The principle aim of this paper is to discuss basic aspects on observing  memory GWs from  SBBHs  using PTAs.
For  comparison, we also analyze inspiral GWs.
The rest of this paper is organized as follows.  In section 2, we provide expressions for the amplitudes of the inspiral and memory GWs.  Then essential properties of PTA noises  are mentioned in section 3.  In section 4,  we study the expected detection numbers for the memory and inspiral GWs with PTAs.
In our simple formulation, we provide various scaling relations that would be useful to discuss PTA observation of GWs from individual SBBHs. Throughout this paper we use the geometrical units with $G=c=1$.

\section{Amplitudes of Inspiral and Memory GWs}

In this section we summarize basic expressions for the amplitudes of the inspiral  and  memory GWs.  For a SBBH (two mass; $m_1$ and $m_2$) in a circular orbit, the amplitudes of the  two polarization modes of inspiral GWs at a frequency $f$ are  given by  the quadrupole formula (see {\it e.g.} Favata 2009a);
\beq
(h_+,h_\times)_{i}=\frac{2\pi^{2/3} \eta M^{5/3} f^{2/3}}D(1+\cos^2I,2\cos I).
\eeq
Here we defined 
 the total mass $M\equiv m_1+m_2$, the reduced mass ratio $\eta=m_1m_2/(m_1+m_2)^2$, the distance $D$, and the inclination angle $I$. 
We have $\eta\le 1/4$ with the equality only for $m_1=m_2$. 
By evaluating the energy loss through gravitational radiation, the time before coalescence is given by
\beq
T_{GW}=\frac{5f^{-8/3}}{256 \pi^{8/3}\eta M^{5/3}}=1.2\times 10^7  \lmk\frac{\eta}{0.25}  \rmk^{-1} \lmk\frac{M}{10^{9.5}M_\odot}  \rmk^{-5/3} \lmk\frac{f}{\rm 1nHz}  \rmk^{-8/3}  {\rm yr}.
\eeq
Around the  optimal frequency for a  PTA, $f\sim 1$nHz, the time  $T_{GW}$ is much longer than a realistic observational period $T_{obs}=O(10)$yr, and thus the inspiral GWs can be regarded as almost periodic signals.

A memory GW is expressed by a net gap of the transverse-traceless components of metric, and it is generated by  anisotropic energy emission from a source.  In the present case of a SBBH,  the relevant  energy emission is due to gravitational radiation, and the energy carried by gravitational radiation  is a nonlinear function of the GW amplitude.  Therefore, the memory GWs of interest are often called the nonlinear GW memory (Christodoulou 1991; Wiseman 
\& Will 1991; Blanchet \& Damour 1992; Thorne 1992; Kennefick 1994; Favata 2009a; 2009b). Since most of GW energy from a SBBH is emitted around the final merger phase with a time duration  $O(M)$, the time profile of a memory wave can be regarded as a step function with a time resolution longer than  $M$.
The gaps of  the two polarization modes for the memory GW are modeled by
\beq
(h_+,h_\times)_{m}=\frac{\eta M g}{384\pi D}\sin^2 I(17+\cos^2 I,0)
\eeq
with an  $O(1)$ parameter $g$  determined by the history of  GW emission, mostly around the merger epoch  (Favata 2009b). In what follows, we use a re-parameterization $g_{12}\equiv g/12$ with $g_{12}\sim1$ from a recent study by Favata (2009b).

{ Next we evaluate the non-dimensional characteristic GW amplitudes defined by $h_c\equiv f \lnk |{\tilde h}_+(f)|^2+|{\tilde h}_\times(f)|^2 \rnk^{1/2}_I$.  Here we denoted the Fourier transformation ${\tilde a}(f)\equiv \int_0^{1/f} a(t) \exp(2i\pi f t)dt$ for a function $a(t)$ in the time domain. We also  introduced the notation $\lnk\cdots\rnk_I$ for the angular averages with respect to  the inclination $I$. } For the inspiral signal, we have
\beq
h_{ci}=\frac{4\pi^{2/3} \eta M^{5/3} f^{2/3} }{\sqrt{5} D}(fT_{obs})^{1/2}=9.2\times 10^{-17} \lmk \frac{\eta}4 \rmk \lmk \frac{M}{10^{9.5} M_\odot} \rmk^{5/3}  \lmk \frac{f}{1 {\rm nHz}} \rmk^{2/3} \lmk \frac{1{\rm Gpc}}{D} \rmk  (fT_{obs})^{1/2}, \label{hci}
\eeq 
where the last factor $(fT_{obs})^{1/2}$ represents the effective signal amplification due to the multiple rotational cycles.  In the same manner, the corresponding amplitude for the nonlinear memory is given by  
\beq
h_{cm}=\frac{g_{12} \eta M }{24\pi^2 D} \sqrt{\frac{1543}{70}}=7.6\times 10^{-16}g_{12} \lmk \frac{\eta}4 \rmk \lmk \frac{M}{10^{9.5} M_\odot} \rmk \lmk \frac{1{\rm Gpc}}{D} \rmk. \label{hcm}
\eeq
Thus, for a given distance $D$,  and an observational frequency $f\sim 1$nHz, the amplitude $h_{cm}$ for the memory GW is larger than the inspiral one $h_{ci}$ in the mass range $M\lsim 10^{10}M_\odot$. But this is not true at frequencies with  $T_{GW}<T_{obs}$, as in the case of a stellar-mass black hole binary  observed by ground based  detectors.  By plugging  $T_{obs}=T_{GW}$ in eq.(\ref{hci}), we obtain $h_{ci}\propto f^{-1/6}$, and  also have $h_{ci}>h_{cm}$ even at the frequency of the last stable orbit (Kennefick 1994; Favata 2009b).

\section{Noises of PTAs}

In this section, we discuss the measurement noise and the background confusion noise for GW observation with a  PTA.   The magnitude of the former, $n_d$, relative to the non-dimensional amplitudes, $h_c$, has the simple frequency dependence $\propto f^{3/2}$ (Rajagopal 
\& Romani 1995). The power index 3/2 can be explained as follows. 
{ When we observe a pulsar, the arrival time  $\tau(t)$ of its pulses is modulated by a passing GW signal $h(t)$ at the Earth as $h(t)\propto {\dot \tau}(t)$.  Therefore, the measurement noise $n_h(t)$ for GW observation is related to the  timing noise $\tau_d(t)$   as $n_h(t)\propto {\dot \tau}_d(t)$. Then, in the frequency domain,  we have $n_d(f)\equiv f \lla |{\tilde n}_h(f)|^2\rra^{1/2}\propto f^2 \lla |{\tilde \tau}_d(f)|^2\rra^{1/2}$. Since the timing noise $\tau_d(t)$ can be regarded as a white noise with $\lla \tau_d(t)\tau_d(t')\rra\propto \delta(t-t')$, we obtain $\lla |{\tilde \tau}_d(f)|^2\rra\propto f^{-1}$, and finally get $n_d(f)\propto f^{3/2}$.}\footnote{ For a given PTA project,  the timing noise level $f\lla |{\tilde \tau}_d(f)|^2\rra^{1/2}$  (essentially corresponding to $\delta t_{rms}(f)$ in Sesana et al. (2009)) is proportional to $f^{1/2}$. In this paper, we examine the expected detection rates of GW signals for individual PTAs, not for a given timing noise level at each frequency (see also  Fig.2 in Sesana et al. (2009)).}

  For a given PTA project, the overall shape of the noise spectrum is characterized by the two parameters $T_{obs}$ and $n_{d0}$. The observational time $T_{obs}$ determines the minimum accessible frequency $f_{obs}\equiv T_{obs}^{-1}$ and the parameter $n_{d0}$ fixes the noise amplitude at a pivot frequency $f_p$ by
\beq
n_{d}(f)=n_{d0} \lmk \frac{f}{f_p}  \rmk^{3/2}.
\eeq
In this paper we take $f_p=10^{-8}$Hz, and use the numerical values  $(T_{obs},n_{d0})=(4.9{\rm yr},5.1\times10^{-15})$ for the  Parks Pulsar Timing Array (PPTA) and $(7.7{\rm yr},1.2\times 10^{-15})$ for the Square Kilometer Array (SKA), extracted from figure 1 in Hobbs et al. (2009) (see also Demorest et al. 2009 for other projects including nanoGrav).


\begin{figure}
\includegraphics[width=100mm]{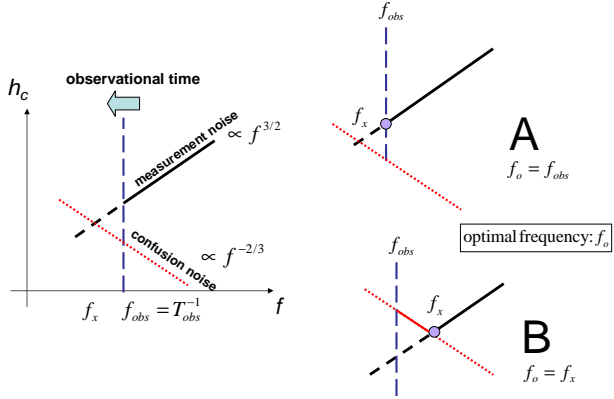}
\caption{ {\it Left panel}: A log-log plot for the relations between the measurement noise $n_d\propto f^{3/2}$ and  the GW background noise $n_b\propto f^{-2/3}$ for detecting a GW with a non-dimensional amplitude $h_c$. The frequency $f_x$ is defined as the intersection of two noises. The lowest observable frequency $f_{obs}=T_{obs}^{-1}$ is determined by the observational time $T_{obs}$. {\it Right panel}: the total PTA noise (solid lines) before (case A) and after (case B) the detection of stochastic GW background. At the optimal frequency $f_o$, the total noise level $n_c$ becomes minimum. We have $f_o=f_{obs}$ for the case A and $f_o=f_{x}$ for the case B. }
\label{fig1}
\end{figure}


Meanwhile the GW background noise $n_b(f)$ by SBBHs has a profile $n_b\propto f^{-2/3}$. { The spectral index $-2/3$ can be  understood with the definition $\Omega_{GW}\propto f^2 n_b^2$ for the normalized energy density  of a GW background per logarithmic frequency interval (see {\it e.g.} Phinney 2001).  In the frequency interval,  the GW energy emitted by a binary is proportional to $f^{2/3}$, as derived with  the Kepler's law. When we sum up GWs from multiple binaries, this power-law profile is unchanged, and we have $\Omega_{GW}\propto f^{2/3}$.  Then we obtain $n_b\propto f^{-2/3}$.}

  We represent the spectrum of the  GW  background noise by
\beq
n_b(f)=n_{bF} r^{1/2} \lmk  \frac{f}{f_p}    \rmk^{-2/3} \label{nb}
\eeq
with a fiducial amplitude $n_{bF}=2.0\times10^{-15}$ and a non-dimensional scaling parameter $r$.  Many theoretical models of structure formation predict  that the GW background $n_b(f)$ is mainly made by the massive end of SBBHs with $M\sim O(10^9M_\odot)$ and the parameter $r$ would be in the range $0.05\lsim r\lsim 20$ (Jaffe  \& Backer 2003; Wyithe \& Loeb 2003; Enoki et al. 2004; Sesana et al. 2008). 	Assuming the characteristic mass parameters $M=10^{9.5}M_\odot$ and $\eta=0.25$, the comoving merger rate $R$ is estimated as\footnote{ While we separately use two parameters $M$ and $\eta$, the chirp mass $\eta^{3/5}M$ is more convenient for the present  argument. Note also that the reduced mass $\eta M$ becomes maximum at $\eta=1/4$ for a given chirp mass.} 
\beq
R=R_F r=3.7\times 10^{-6} r {\rm Gpc^{-3} yr^{-1}}\label{rate}
\eeq
(from an expression in  Phinney 2001), 
 corresponding to the total merger rate on our past light-cone ${\cal R}_T\sim 2.2\times 10^{-3} r ~{\rm yr^{-1}}$. Here we neglected redshift dependence of the comoving rate $R$.

From a continuity equation in the Fourier space, the frequency distribution  of the SBBHs per comoving volume is given by $\frac{dn}{d\ln f}=R \frac{dt}{df}f=\frac83 RT_{GW}$.  This expression will be used in the next section.  In the same  manner, the total number of the SBBHs around a  frequency $f$ within the observational bandwidth $f_{obs}$ is given  by $8{\cal R}_TT_{GW}f_{obs}/(3f)$, and is much larger than unity at $f\lsim 10^{-8}$Hz for a plausible value of $r$.  This means that an inspiral GW signal from a binary at a typical cosmological distance would be buried under the smooth GW background. At $f\gsim 10^{-8}$Hz, discreteness effects of SBBHs might show up for the background (Sesana et al. 2008).  But, for the fiducial value $n_{bF}$,  we have $n_d\gsim n_b$  for PPTA or SKA in the high frequency regime, and,  in this paper,  we simply neglect the discreetness effect of the background.

Now we discuss the total noise spectrum $n_c(f)=\max[n_d(f),n_b(f)]$ made by the measurement noise $n_d(f)$ and the confusion noise $n_b(f)$ (see figure 1).  The characteristic frequency $f_x$ for their intersection is solved as 
$
f_x=f_p n_{bF}^{6/13} n_{d0}^{-6/13} r^{3/13}$ and  $n_c(f_x)=n_{bF}^{9/13} n_{d0}^{4/13} r^{9/26}$. For an observational time $T_{obs}$ shorter than $f_x^{-1}$, the total noise $n_c(f)$ is determined mainly  by the measurement noise as $n_c(f)=n_d(f)$ (case A in figure 1). But, if the time $T_{obs}$ is longer than $f_x^{-1}$ and the detection of the  GW background is within reach, the effective noise $n_c(f)$ is  a piecewise power-law function (case B in figure 1).
For both cases, the optimal frequency associated with the minimum value of the noise level $n_c(f)$ is given by $f_o=\max[f_x,f_{obs}]$.

\section{Expected Numbers of Detections}

In this section, we estimate how many detections we can expect for  inspiral and memory GWs of  nearby SBBHs with PTAs.  The signal-to-noise ratio for the detection is given by 
$
SN={h_c}/{n_c}
$ 
from which we can inversely obtain the observable comoving distances $D$ as well as the observational volumes $4\pi D^3/3$ for a given threshold $SN$. Here, to deal with these geometrical quantities, cosmological effects can be safely neglected, since the detectable binaries would be at a relatively low redshift.

Then, using the relevant comoving number densities (${dn}/{d\ln f}$ for the inspiral signals and $RT_{obs}$ for the memory signals), we obtain the expected numbers of detectable events  in a logarithmic frequency interval as
\beq
\frac{dN_i}{d\ln f}=\frac{2 M^{10/3} R f^{5/6} T_{obs}^{3/2} \eta^2}{ 5^{1/2} 3 \pi^{2/3} n_c(f)^3 SN^3},~~~
\frac{dN_m}{d\ln f}=\frac{1543^{3/2} g_{12}^3 M^3  R  T_{obs} \eta^3}{967680 ~70^{1/2} \pi^6 n_c(f)^3 SN^3}  \label{rate1}
\eeq
for the inspiral (i) and the memory (m) GWs.
Since we have $n_d\propto f^{3/2}$ and $n_b\propto f^{-2/3}$, the total detection rates $N_i$ and $N_m$ are dominated by signals around the optimal frequency $f_o$ and we simply put  
$N_i= \lmk {dN_i}/{d\ln f}\rmk_{f_o}$, and $N_m =\lmk {dN_m}/{d\ln f}\rmk_{f_o}$ for our order-of-magnitude estimation.

So far we have only considered SBBHs  at their massive end $M\sim 10^{9.5}M_\odot$.  But, here,  we briefly comment on  the mass dependence of the detectable binaries.  For a given  merger rate $dR/d\ln M$ per logarithmic mass interval, the numbers of detectable binaries are proportional to $M^\alpha dR/d\ln M$ with $\alpha=10/3$ for inspiral and $\alpha=3$ for the memory signals (see eqs.(\ref{rate1})).  But theoretical studies (see {\it e.g.} Sesana et al. 2009) predict that the mass distribution $dR/d\ln M$ of the merger rate would be apparently less steep than $\propto M^{-3}$ in the mass range lower than $M\sim 10^{9}M_\odot$ ({\it e.g.} at $M=10^8M_\odot$). Thus the detectable GW signals would be mainly made by the massive end with $M\gsim 10^9M_\odot$ and our basic prescription so far would be justified.  In the following demonstration, we take $M= 10^{9.5} M_\odot$ and $\eta=0.25$.

Now let us discuss the prospects for detecting  inspiral and memory GWs with the two representative projects; the ongoing PPTA and the planned SKA.  For the fiducial background at $r=1$,  we have the optimal point $(f_o,n_c(f_o))=(6.5 \times 10^{-9}{\rm Hz}, 2.7\times 10^{-15})$ with PPTA and $(1.3 \times 10^{-8}{\rm Hz}, 1.7\times 10^{-15})$ with SKA. In figure 2, we plot the expected event numbers as  functions of the scaling  parameter $r$ defined relative to the fiducial background model (see eqs.(\ref{nb}) and (\ref{rate})). In this figure we set the threshold at $SN=1$, since its dependence is straightforward.  The numbers $N_i$ and $N_m$ change their power-law indexes at the transition point $r_T\propto T_{obs}^{-13/3} n_{d0}^2$ where the coincidence $f_x=f_{obs}$ occurs for a given project. For PPTA the transition point is at $r_T=1$, since we have $f_x=f_{obs}$ for the fiducial background level (consistent with figure 1 in Hobbs et al. 2009). But the point $r_T$ becomes smaller for SKA whose observational period is longer with a smaller measurement noise.  At $r<r_T$, we have $f_o=f_{obs}$ and  the dependencies on the rate $r$ are very simple 
$
N_i\propto r T_{obs}^{31/6}$ and $ N_m\propto r T_{obs}^{11/2}
$,
 as the background noise is not important in this regime (case A in figure 1).  At $r>r_T$, we have
$
N_i\propto r^{2/13} T_{obs}^{3/2}$ and $ N_m\propto r^{-1/26} T_{obs}
$,  and these two numbers depend very weakly on the parameter $r$. 
Interestingly, the number $N_m$ is now  a decreasing function of $r$.

\begin{figure}
\includegraphics[width=84mm]{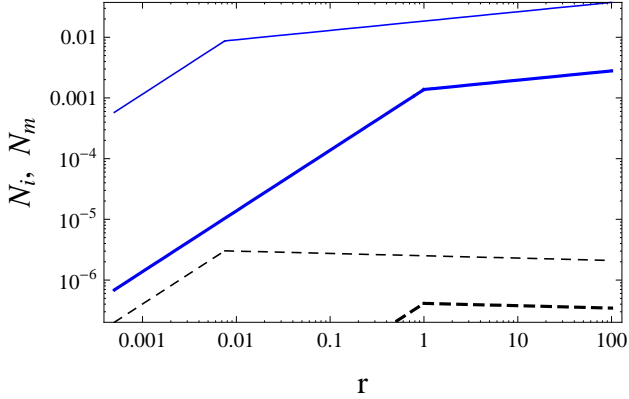}
\caption{The expected numbers of events $N_i$ (solid curves) and $N_m$ (dashed curves) with $SN\ge 1$ for the reference value  $M=10^{9.5}M_\odot$.  The horizontal axis represents the  strength $r$ of the GW background  relative to the fiducial value $n_{bF}$. The thick curves are for PPTA with the transition point at $r_T=1$ and the thin curves are for SKA with $r_T=7.5\times 10^{-3}$. We have the scaling relations $N_i,N_m\propto r$ at $r<r_T$, and $N_i\propto r^{2/13}$ and $N_m\propto r^{-1/26}$ at $r>r_T$.}
\label{fig2}
\end{figure}

For each PTA project, the expected number $N_m$ for the memory signals is 3-4 orders of magnitude smaller than  the number $N_i$ for the inspiral signals.  { As we discussed before, the comoving number densities for the merger  and the inspiral signals per logarithmic frequency interval are given by $RT_{obs}$ and $8RT_{GW}/3$ respectively. In the frequency regime relevant for PTAs, the former is much smaller than the latter, since we have $T_{GW}\gg T_{obs}$. Therefore, even though the detectable distance is larger for the merger signals as indicated by eqs.(4) and (5), their detection rate becomes smaller than that of the inspiral signals. }

Figure 2 also shows that SKA would have at least $\sim 10$ times larger events $N_i$ than PPTA, but is still unlikely to detect an inspiral event during its operation period. These qualitative predictions for circular SBBHs  would be fairly robust, and would not be changed with a more detailed analysis ({\it e.g.} using an elaborate mass function of SBBHs  around $M\sim 10^9 M_\odot$).

The author is grateful to T. Tanaka for carefully reading the manuscript.  He also thanks the referee for valuable comments to improve the draft.
This work was supported by the  Grants-in-Aid for Scientific Research 20740151 from the Ministry of Education, Culture, Sports, Science and Technology (MEXT) of Japan.

{ After submission of this paper, there appeared  related studies on PTAs by Pshirkov et al. (2009), and van Haasteren and Levin (2009). They predicted  higher rates for detection of the memory GWs.  The difference is mainly due to the treatment of the background  noise. Without the background noise,  the detection rates are obtained by extrapolating the lines at $r<r_T$ in Fig. 2 to the regime $r>r_T$.}


\end{document}